# Phosphorous Alloying: Controlling the Magnetic Anisotropy in Ferromagnetic GaMnAsP Layers

M.Cubukcu[1], H.J. von Bardeleben[1], Kh.Khazen[1], J.L. Cantin[1], O.Mauguin[2], L. Largeau[2], and A.Lemaître[2]

[1]*Institut des Nanosciences de Paris*
*Université Pierre et Marie Curie, UMR 7588 au CNRS*
*140 rue de Lourmel, Paris, F-75015 Paris, France*
[2]*Laboratoire de Photonique et de Nanostructures, CNRS*
*Route de Nozay, 91460 Marcoussis, France*

**Abstract :**

Phosphorous alloying of GaMnAs thin films has been used for the manipulation of the magnetic anisotropies in ferromagnetic $Ga_{0.93}Mn_{0.07}As_{1-y}P_y$ layers. We have determined the anisotropy constants as a function of temperature for phosphorous alloying levels between 0 and 8.8 at % for a Mn doping level of ≈ 7at%. We show that it is possible to obtain layers with robust ferromagnetism and either in-plane or out-of plane easy axes with small barriers for magnetization reorientation by phosphorous alloying with y≤ 6at% or y≥ 6at%. The critical temperatures are not significantly increased by the P alloying.

Diluted magnetic semiconductor GaMnAs thin films doped with Mn at concentration between 5 to 10% have shown their high potential for spintronics application and their interesting micromagnetic properties which are closely related to the valence band structure and the hole concentration. Generally, these films grown epitaxially on GaAs substrates are characterized by strong magnetocrystalline anisotropy due to strain induced by the lattice mismatch between the GaMnAs epitaxial film and the underlying substrate [1-4]. For device application it is interesting to control the magnetic anisotropy and create conditions for easy switching of the magnetization by applied fields or induced by currents. Various techniques have been proposed to manipulate this process independently from the Mn doping concentration. Examples are carrier depletion in diode structures by the application of an external electric field [5] or the use of multilayers structures composed of GaMnAs and piezoelectric [6] or ferroelectric top layers [7]. However, all these approaches give rise to serious material problems and require multiple ex-situ growth processes. It would be extremely interesting to overcome this bottleneck and very recently it has been shown that additional alloying with phosphorous allows modifying the epitaxial strain [8, 9]. In this letter, we report the results of an extensive study of the anisotropies in GaMnAsP quaternary thin films which show that it is possible to combine high Mn doping levels required for robust ferromagnetism with small barriers for magnetization switching obtained by adequate phosphorous alloying.

50 nm thick $Ga_{1-x}Mn_xAs_{1-y}P_y$ films were grown by low temperature molecular beam epitaxy on GaAs (100) substrates. The concentration of phosphorous was varied from y=0 to y=0.08 and the Mn concentration was set to x≈ 0.07. After the growth the samples were thermally annealed under $N_2$ atmosphere at 250° for 1 hour. The magnetic properties of the films were characterized by magnetometry using a superconducting quantum interference device (SQUID) and ferromagnetic resonance spectroscopy (FMR). The uniaxial strain $\varepsilon_{zz}$ was determined by high resolution X-ray diffraction. The magnetocrystalline anisotropy constants and associated anisotropy fields were determined by X-band (9 GHz) ferromagnetic resonance (FMR) spectrometry. The angular dependence of the $Ga_{1-x}Mn_xAs_{1-y}P_y$ FMR spectra were measured in two planes, (001) and (110). These two sets of measurements enable us to determine the resonance position for the high symmetry orientation of $Ga_{1-x}Mn_xAs_{1-y}P_y$ film: H//[001],[110],[1-10] and [100] from which the anisotropy constants can be determined [1,10].

The sheet resistivity of the series of $Ga_{1-x}Mn_xAs_{1-y}P_y$ samples containing different concentrations of P is shown in Fig.1. The samples were taken from different growth runs with very similar Mn concentrations. They all present very similar critical temperatures between 110K and 135K. We see that the conductivity of the samples is metallic in all cases as expected for a Mn doping of $x \approx 0.07$ but varies within a factor of three. The weak dependence of the Tc ($\sim p^{1/3}$) which is predicted by the mean field model [11,12] explains that the Curie temperature in this series varies nevertheless only slightly. In principle we should expect a systematic variation of the hole concentration with the phosphorous content as the acceptor level becomes increasingly deeper. However, the Mn incorporation is also influenced by the P alloying and more detailed measurements are required in order to investigate this interdependence.

Fig.2(a) shows typical FMR spectra of the $Ga_{1-x}Mn_xAs_{1-y}P_y$ samples with concentrations of P y=0, 0.056 and 0.088. The applied magnetic field is in each case aligned parallel to the respective easy axes. For phosphorous concentrations y=0 to y=0.056 the films have an easy in-plane axis along [100], with resonance fields near H=1632 Oe (1), H=2831 Oe (2) and hard axes along the [001] direction H=7816 Oe, H=3761 (not shown here). This corresponds to the case of compressively strained films on GaAs substrates. For higher concentrations of phosphorus we observe a change of the easy axis of magnetization to [001] i.e. perpendicular to the film plane. The dominant uniaxial anisotropy constant $K_{2\perp}$ can be continuously varied from highly negative to highly positive values by adjusting the P alloying level. The FMR spectra of the highly P doped samples show in addition to the uniform mode (fig.2a) two spin wave resonances with resonance fields near H=1289 Oe, H=266 Oe respectively. The hard axis of the sample with y=0.088 was observed to be along the [110] direction, H=4977 Oe (not shown here).

Analyzing the FMR spectra via the Smit-Beljers formalism [13] and taking into account the minimization of the free energy for the equilibrium states of the magnetization [1], we have determined the four magnetic anisotropy constants of the samples. As it has been shown previously, the dominant constant for GaMnAs layers on GaAs or GaInAs substrates with typical Mn concentrations of x=0.07 is $K_{2\perp}$ which is proportional to the uniaxial strain induced by the lattice mismatch. Its value at low temperature and its variation with temperature are shown in fig.2 (b) for three samples representing the cases of high tensile, high compressive and close to zero strain. These cases correspond to P concentrations of y=0 0, 0.088 and 0.056. For the y=0 sample the $K_{2\perp}$ constant at T=4K has a negative value of $-6.10^4$ erg/cm$^3$; with increasing P doping level its value increases monotonously and passes though zero (for y=0.056). For this range of phosphorous concentration the four anisotropy constants are small and of comparable numerical value. With further increase of the [P] $K_{2\perp}$ becomes positive and the easy axis of magnetization switches to out-of-plane. In figure 2 (c), we show the anisotropy fields 2K/M as a function of phosphorous concentration. The fact of dividing the anisotropy constant by the magnetization allows us to correct the variation of $K_{2\perp}(y)$ for potentially slightly different Mn concentrations. In a previous publication we have experimentally confirmed that the $K_{2\perp}$ constant depends linearly on the carrier concentration and the uniaxial strain. Assuming that the mobility is similar for all samples of this series we can take into account slightly different hole concentrations from the conductivity values and we then expect a linear variation of the anisotropy fields normalized to a constant hole concentration with the uniaxial strain. Indeed, this is observed in the fig.2 (d). In this figure the anisotropy field $H_{2\perp}$ at 4K and the its normalized value ($H_{2\perp} \cdot \rho xx$) are plotted as a function of the uniaxial deformation $\varepsilon_{zz}$. The linear variation is in good agreement with the mean field model, as it has been shown also for the case of GaMnAs [10]. Our results demonstrate that it is possible to manipulate the uniaxial anisotropy field to any desired value just by varying the P doping level without any loss of sample quality.

It can be seen in fig.2(d) that the linear fit to these results does not pass through the origin. Whereas such behaviour would have been expected for the total perpendicular anisotropy field due to the contribution of the demagnetization field this is not the case for the $H_{2\perp}$ defined by $2K_{2\perp}/M$. We describe this small deviation to the fact that the samples have also slightly different values of the exchange splitting parameters due to the alloying which will scatter the corresponding $K_{2\perp}$ values [3].

The observation of spinwave resonances in addition to the uniform mode spectrum in highly P doped samples gives us important information on the physics in these quaternary alloys. It allows us to determine the effective exchange integral between the Mn ions; according to theoretical predictions its value is expected to be increased relative to the case of GaMnAs [14]. In order to deduce this parameter we have applied the phenomenological model proposed by Liu et al. for [001] and [100] directions of the applied field [15]. The exchange constant, D, deduced at T=40K is ~2.733 Tnm$^2$ which by considering the magnetization value at this temperature leads to the spin stiffness, A=0.06 pJ/m. This value is slightly larger than what has been observed in GaMnAs/GaInAs (x=0.07) samples via magneto-optical Kerr effect microscopy [16]; from this value we deduce an effective exchange integral between the Mn ions of $J_{Mn-Mn}$=0.22meV. Note that in contradiction to theoretical predictions the effective exchange integral has not increased considerably by the P alloying.

FMR measurements have been equally performed at a second microwave frequency Q-Band (35GHz) in order to determine the Gilbert damping factor from the variation of the uniform mode line width with the microwave frequency. As shown in the inset of fig.2 the damping factor of the sample with highest P concentration is $\alpha=0.012$ at low temperature; this value is about three times higher than the one reported previously for phosphorous free GaMnAs layers[17]. The damping factor varies only weakly with temperature between 4K and $0.8T_C$ and increases as is usually observed when the temperature approaches $T_C$.

A further interesting aspect is the small value of the inhomogeneous part of the FMR linewidth. Its value is only 40 Oe which is comparable to the best state of the art GaMnAs samples grown on GaAs(001) substrates. It is however an order of magnitude smaller than that reported in the case of GaMnAs layers with perpendicular anisotropy, which had been obtained by epitaxial growth on GaInAs substrates (the tensile strained samples). The inhomogeneous linewidth is a measure of the magnetic homogeneity of such films and its low value in phosphorous alloyed films manifests the absence of dislocation related inhomogeneous dopant and charge carrier distributions which are a major drawback for the use of GaInAs buffer layers.

From the numerical values of the anisotropy constants we can calculate the free energy density E as a function of the crystalline orientations $(\theta,\phi)$. In fig.3 (a), (b), (c) we show such surfaces for zero the applied magnetic field for the three prototype samples with y=0, 0.056, 0.088 respectively. These surfaces show the energy barriers for the magnetization reorientation and define the preferential axes of the magnetization. A more detailed view can be obtained from the cross sections of the surface in particular planes. For example in fig.3 (d) we have plotted the out-of-plane angular variation of the free energy density as a function of $\theta$ angle. These figures illustrates that while for the two extreme cases the difference between the energy along the easy and hard axes are considerably large, for the intermediate case of the sample with y=0.056 the variation of the energy becomes very small.

This is a very important result as it indicates that one can reduce the energy barriers of the magnetization switching for both out of plane and in-plane easy axis configurations. This makes such layers interesting objects in FM/NM/FM trilayer structures currently studied for their tunnelling magneto resistance effects.

The energy surfaces for the in-plane orientations ($\theta$=90) of the magnetization for samples with y=0 and 0.056 are shown in fig.3 (e) and (f) respectively. At low temperatures the easy axis of the magnetization in the absence of applied magnetic fields is not along any high symmetry direction of the film due to competing second order and forth order anisotropy fields. They are oriented along the intermediate direction $\varphi$ =-18° and -19.8° with respect to the [100] direction for y=0 and y=0.056 respectively.

When we raise the temperatures (to T= 20K and T=40K respectively) the easy axis monotonously shifts to the [1-10] direction. These particular orientations correspond to the predominance of the $K_{2//}$ and $K_{4//}$ constants, as shown in fig.3 (g). Thus the temperature is another controllable parameter in addition to the strain which can be used to manipulate the orientation of the magnetization of $Ga_{1-x}Mn_xAs_{1-y}P_y$ ferromagnetic films with a fixed Mn concentration [10 ,18].

In conclusion we have shown that ferromagnetic phosphorous alloyed $Ga_{1-x}Mn_xAs_{1-y}P_y$ layers with x≈0.07 and y=0 to 0.088 of high structural and magnetic qualities can be grown by low temperature molecular beam epitaxy. The conductivity, critical temperature and damping factor are only weakly influenced by the phosphorous alloying. Choosing phosphorous doping levels around y=0.06 layers with in plane or out of plane easy axes with small barriers for magnetization switching can be obtained. Such layers are promising objects for application in spintronics devices.

**Acknowledgments**: We thank A.Maitre from LPN laboratory for the electrical transport measurements and T.Cren (INSP) for his contribution to the energy surface representations.

# Figure Captions

**Fig. 1(colour on line)**: Electrical resistivity of the different $Ga_{1-x}Mn_xAs_{1-y}P_y$ samples as a function of temperature. The arrows indicate the approximate Curie temperatures.

**Fig. 2(colour on line)**: (**a**) FMR spectra of $Ga_{1-x}Mn_xAs_{1-y}P_y$ 50 nm thick annealed samples for applied fields along the easy axis; the easy axis of magnetization is parallel to [100] for the samples y=0 (**1, black**), y=0.056 (**2,blue**) and has switched to out-of-plane, i.e. //[001], for the sample y=0.088 (**3,red**). Inset: Damping factor as a function of temperature and magnetic field orientation; H//[001] squares (black), H//[110] up triangles(green), H//[100] circles(red), H//[1-10] down triangles (blue). (**b**) The magnetocrystalline anisotropy constants as a function of temperature for the samples y=0 (blue), y=0.056(black) and y=0.088(red). The lines are guides for the eyes. The magnetic anisotropy fields at 4K as a function the phosphorous concentration (**c**) and the anisotropy fields $H_{2\perp}$ at 4K before (black) and after normalization (red) for a constant hole concentration as a function of strain $\varepsilon_{zz}$ (**d**).

**Fig. 3(colour on line)**: 3D plot of free energy densities for $Ga_{1-x}Mn_xAs_{1-y}P_y$ at 4K with y=0 (**a**), y=0.056 (**b**) and y=0.088 (**c**). The energy density with y=0 (black), y=0.056 (blue) and y=0.088 (red) as a function of the out of plane orientation $\theta$, $\varphi = 45°$ (**d**). Free energy densities as a function of the in-plane magnetization orientation $M = |M|\|\cos\varphi, \sin\varphi, 0|$ with y=0 at T=4K (black) and T=20K (red) (**e**) and y=0.056 at T=4K (black), 20K (red), 40K (blue) (**f**). The directions of the energy minima rotate toward [1-10] with increasing temperature. The magneto crystalline anisotropy constants $K_{2//}$ and $K_{4//}$ as a function of temperature for the samples y=0, y=0.056 and y=0.088 (**g**). The lines are guides for the eyes. The change of easy axes is indicated by the circles.

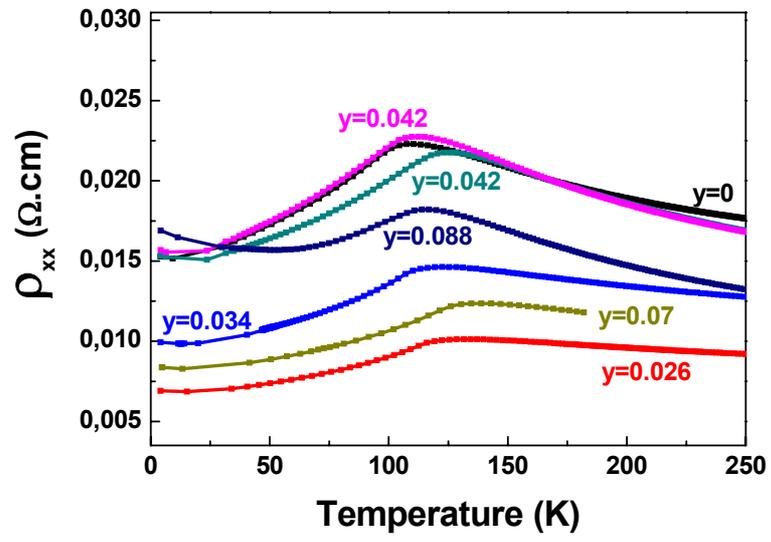

Figure 1

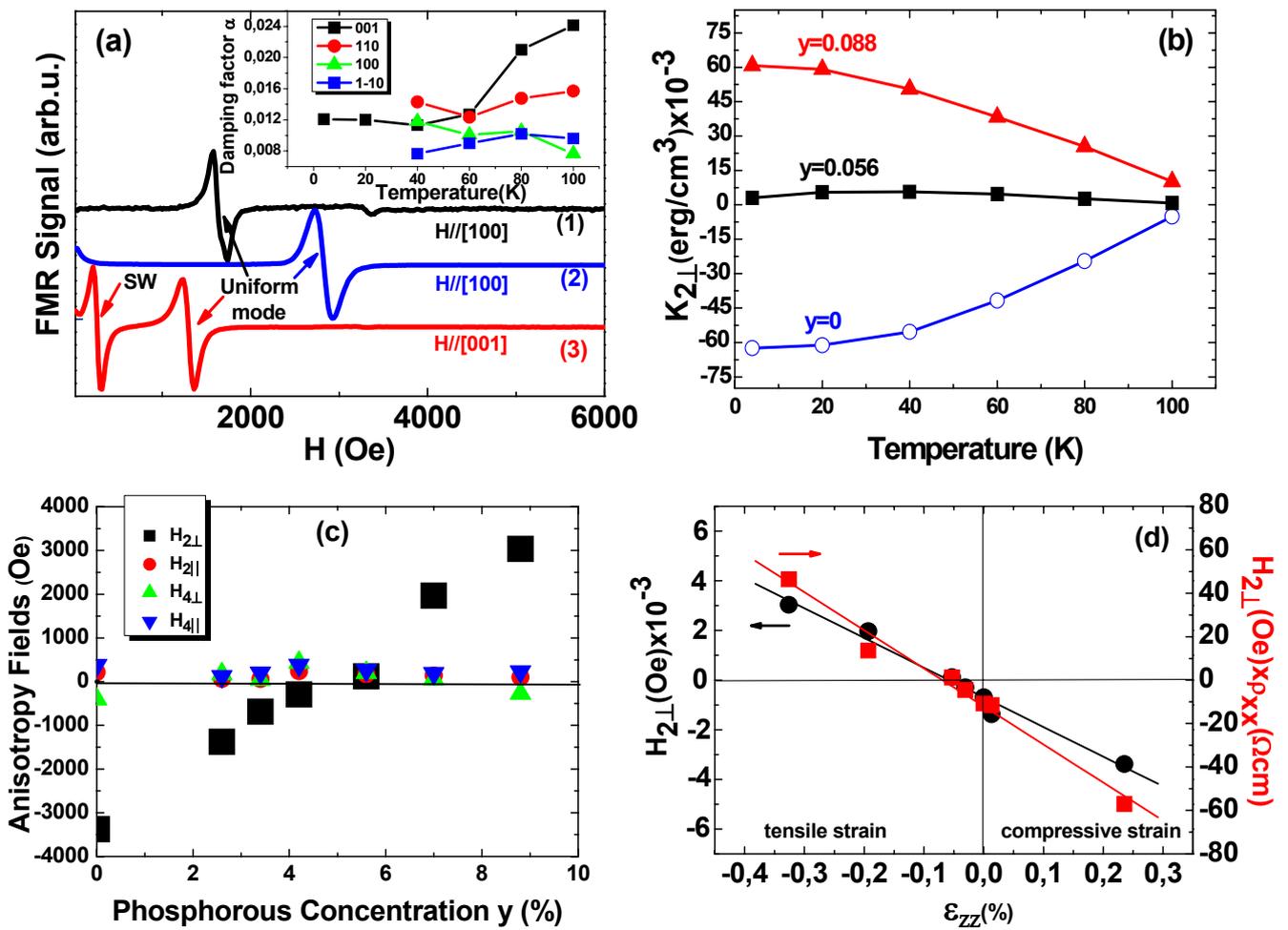

**Figures 2a, 2b, 2c, 2d**

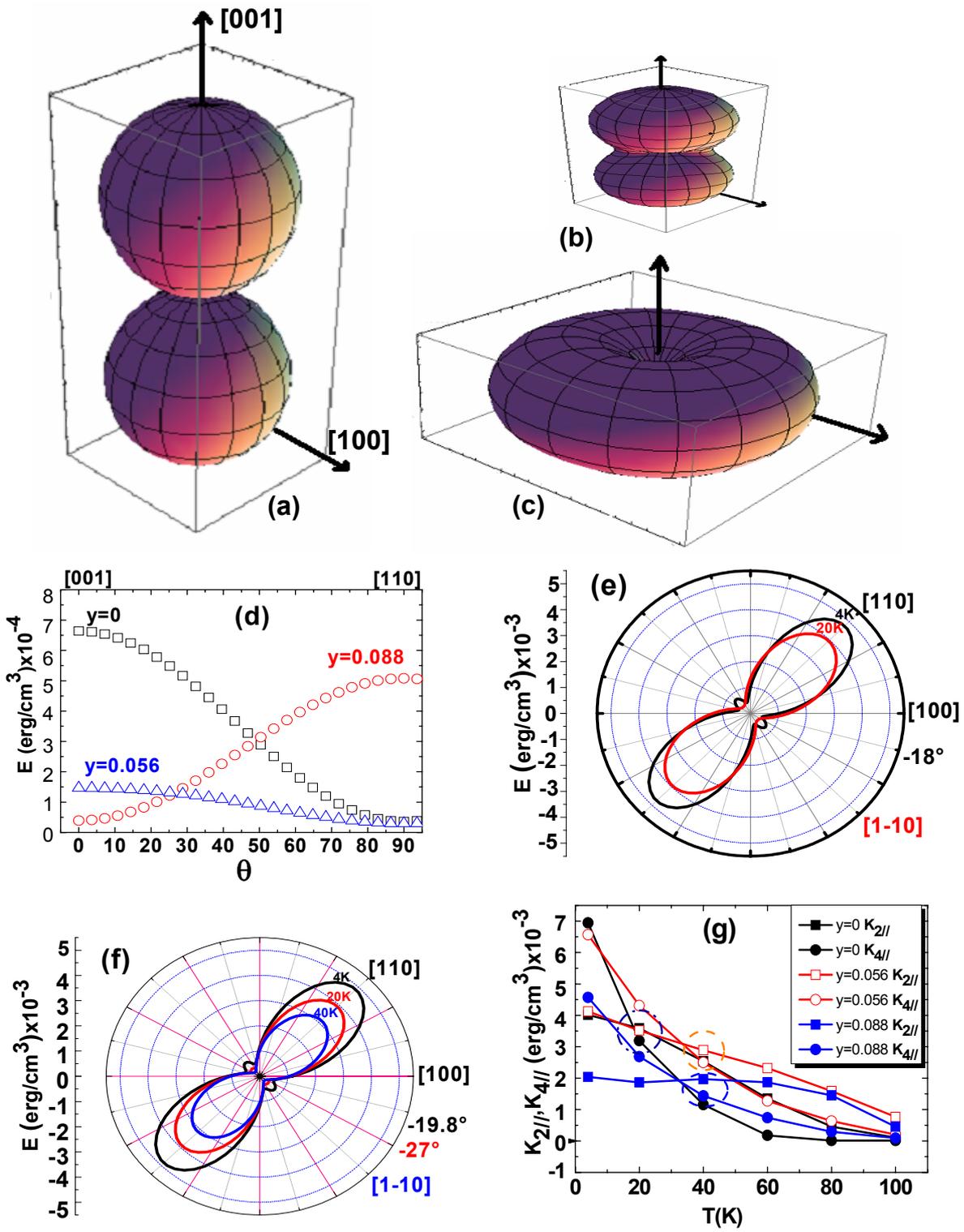

**Figures 3a, 3b, 3c, 3d, 3e, 3f, 3g**